\documentclass[preprint,12pt]{aastex}%
\usepackage{amssymb}
\usepackage{amsmath}
\usepackage{amsfonts}
\usepackage{graphicx}%
\begin{document}

\title{Chameleon gravity and satellite geodesy}
\author{J.R. Morris \bigskip\altaffilmark{}}

\begin{abstract}
We consider the possibility of the detection of a chameleon effect by an earth
orbiting satellite such as LAGEOS, and possible constraints that might be
placed on chameleon model parameters. Approximate constraints presented here
result from using a simple monopole approximation for the gravitational field
of the earth, along with results from the Khoury-Weltman chameleon model,
solar system constraints obtained from the Cassini mission, and parameter
bounds obtained from the LAGEOS satellite. It is furthermore suggested that a
comparison of ground-based and space-based multipole moments of the
geopotential could reveal a possible chameleon effect.

\end{abstract}


\affil{Physics Dept., Indiana University Northwest,}\affil{3400 Broadway, Gary, Indiana 46408 USA}


\bigskip









\section{Introduction}

Newtonian gravity, a Newtonian limit (weak fields, nonrelativistic motions) of
general relativity, has been successful in describing gravitational phenomena
in laboratory settings. However, models of modified gravity have been proposed
which may give rise to small corrections to Newtonian gravity, perhaps
manifesting themselves on different distance scales or under certain
environmental influences. One such interesting modification is the
\textquotedblleft chameleon gravity\textquotedblright\ model
\citep{KW PRD,KW PRL}, wherein the effective coupling of gravitation to matter
can depend upon the environment. In particular, in regions of high mass
density, the scalar component of the interaction (the scalar
\textit{chameleon} field, $\phi$) develops a large mass and a short range, so
that its effect is suppressed, whereas in regions of low mass density the
field can become effectively massless and long ranged, and consequently have a
much more pronounced effect, reflected in a different value of the effective
gravitational constant. Therefore, according to the chameleon model,
earth-based gravity is different from space-based gravity.

\bigskip

Attention is focused here on the motion of a satellite, mimicking a
\textquotedblleft test particle\textquotedblright. Assuming the chameleon
gravity hypothesis, a chameleonic acceleration of the satellite results, due
to a departure from pure geodesic motion. Therefore, sensitive measurements of
any such departure can serve to set bounds on chameleon model parameters.
Here, we use the reported bound on a fifth force coupling constant $\alpha$,
obtained from a data analysis of the LAGEOS satellite, in conjunction with
results from the Khoury-Weltman chameleon model, along with solar system
constraints from the Cassini mission to establish possible bounds on the
chameleon model. Numerical estimates are made, using a simple monopole model,
as in \citep{KW PRD,KW PRL}. These estimates lead us to establish approximate
constraints on the chameleonic acceleration $a_{\phi}$, and constraints on the
chameleon coupling parameter $\beta$ may be approximated, which describes the
strength with which the chameleon field $\phi$ couples to matter. Furthermore,
we suggest that a comparison of ground-based and space-based multipole moments
of the geopotential could provide evidence for the existence or nonexistence
of a measurable chameleon effect.

\section{The chameleon model}

The chameleon model can be considered to arise from a scalar-tensor theory
where the scalar field $\phi$ couples to the Ricci scalar $\tilde{R}[\tilde
{g}_{\mu\nu}]$ in the Jordan frame representation with metric $\tilde{g}%
_{\mu\nu}$. A conformal transformation to the Einstein frame representation,
with metric $g_{\mu\nu}$, removes the coupling of $\phi$ from the curvature,
but a coupling to the matter sector emerges. The action is given by%
\begin{equation}
S=\int d^{4}x\sqrt{g}\left\{  \frac{1}{2\kappa^{2}}R[g_{\mu\nu}]+\frac{1}%
{2}g^{\mu\nu}\partial_{\mu}\phi\partial_{\nu}\phi-V(\phi)\right\}
+S_{m}\left[  A^{2}(\phi)g_{\mu\nu},\psi\right]  \label{e1}%
\end{equation}

where the Jordan frame metric $\tilde{g}_{\mu\nu}$ and Einstein frame metric
$g_{\mu\nu}$ are related by the conformal transformation $\tilde{g}_{\mu\nu
}=A^{2}g_{\mu\nu}$ and $g=|\det g_{\mu\nu}|$. Matter fields are represented
collectively by $\psi$ and $\kappa^{2}=8\pi G$. The matter part of the action
is%
\begin{equation}
S_{m}=\int d^{4}x\sqrt{\tilde{g}}\mathcal{\tilde{L}}_{m}(\tilde{g}_{\mu\nu
},\psi)=\int d^{4}x\sqrt{g}\mathcal{L}_{m}[A^{2}(\phi)g_{\mu\nu},\psi]
\label{e2}%
\end{equation}

The function $A(\phi)=\exp(\beta\kappa\phi)$ is an increasing function of
$\phi$ with the constant $\beta$ representing the strength of the coupling of
$\phi$ to matter. The potential $V(\phi)$ is taken to be a decreasing function
of $\phi$, and a variation of the action $S$ with respect to $\phi$ leads to
an effective potential in the Newtonian limit given by%
\begin{equation}
V_{eff}(\phi)=V(\phi)+\bar{\rho}A(\phi)=V(\phi)+\bar{\rho}e^{\beta\kappa\phi}
\label{e3}%
\end{equation}

where $\bar{\rho}$ is a $\phi$ independent conserved energy density of
nonrelativistic matter in the Einstein frame. The action $S$ yields the
equations of motion%
\begin{equation}%
\begin{array}
[c]{c}%
R_{\mu\nu}-\frac{1}{2}g_{\mu\nu}R=-\kappa^{2}T_{\mu\nu}=-\kappa^{2}\left[
T_{\mu\nu}^{(\phi)}+T_{\mu\nu}^{(m)}\right] \\
\square\phi+\dfrac{\partial V}{\partial\phi}-\dfrac{\partial\mathcal{L}_{m}%
}{\partial\phi}=0\\
\dfrac{du^{\nu}}{ds}+\Gamma_{\alpha\beta}^{\nu}u^{\alpha}u^{\beta}%
-(\partial_{\mu}\ln A(\phi))\left[  g^{\mu\nu}-u^{\mu}u^{\nu}\right]  =0
\end{array}
\label{e4}%
\end{equation}

We use a metric with negative signature, $(+,-,-,-)$. The line element is
$ds^{2}=g_{\mu\nu}dx^{\mu}dx^{\nu}$, the velocity of a test particle is
$u^{\alpha}=dx^{\alpha}/ds$, $\square\phi=\nabla_{\mu}\partial^{\mu}\phi$ with
$\nabla_{\mu}$ the covariant derivative, and $T_{\mu\nu}$ is built from a
chameleon part $T_{\mu\nu}^{(\phi)}$ and a matter part $T_{\mu\nu}^{(m)}$,
with%
\begin{equation}
T_{\mu\nu}=\frac{2}{\sqrt{g}}\frac{\partial(\sqrt{g}\mathcal{L})}{\partial
g^{\mu\nu}},\ \ \ \ \mathcal{L}=\frac{1}{2}(\partial\phi)^{2}-V(\phi
)+\mathcal{L}_{m} \label{e5}%
\end{equation}

We will be interested in the Newtonian limit of the above model, in which case
the equations of motion reduce to
\begin{subequations}
\label{1}%
\begin{align}
\nabla^{2}\Phi &  =\kappa^{2}(T_{00}-\frac{1}{2}T_{\lambda}^{\lambda
})\smallskip=\kappa^{2}\left[  \frac{1}{2}\bar{\rho}A(\phi)-V(\phi)\right]
\label{e6a}\\
\nabla^{2}\phi &  =\dfrac{\partial V}{\partial\phi}+\beta\kappa\bar{\rho
}A(\phi)\label{e6b}\\
\vec{a}  &  =\dfrac{d^{2}\vec{x}}{dt^{2}}=-\nabla\Phi-\nabla(\ln A)
\label{e6c}%
\end{align}

where $\Phi=\frac{1}{2}h_{00}$ is the Newtonian gravitational potential, and
we will assume no back reaction on $\Phi$ due to $\phi$. [The terms on the
right hand side of (\ref{e6a}) are from the stress-energy tensor, which is
generated by both matter and the chameleon field. Outside of a mass
distribution the matter density $\bar{\rho}\approx0$ and the contribution to
$(T_{00}-\frac{1}{2}T_{\lambda}^{\lambda})$ is due to the chameleon, and
reduces to $(T_{00}^{(\phi)}-\frac{1}{2}T_{\ \ \ \lambda}^{(\phi)\lambda
})=-V(\phi)$ for $r\gg r_{S}$, where $r_{S}$ is the Schwarzschild radius of
the source. This term $V(\phi)$ is expected to be negligible in a low density
region far from the source, where $\phi$ can become much larger than in a high
density region (with $\kappa\phi\lesssim1$).]

\bigskip

Let us note here that in the Newtonian limit the equation of motion for the
scalar field can be written as $\square\phi+\partial V_{eff}/\partial\phi=0$
where $V_{eff}$ takes the form given in (\ref{e3}).
[Refs.\citep{KW PRD,KW PRL} can be consulted for further details.] Now
consider the potential $V(\phi)$ to be given by $V=M^{5}/\phi$, or perhaps
more generally, by $V=M^{4+n}/\phi^{n}$. The functions $V(\phi)$ and $A(\phi)$
in the effective potential $V_{eff}$ have competing behaviors, so that
$V_{eff}$ exhibits a minimum at some value $\phi_{\text{min}}$ which will
depend upon the local matter density $\bar{\rho}$, as seen from (\ref{e3}).
The mass of the scalar field $\phi$ is computed from $V_{eff}$ where the
curvature $V_{eff}^{\prime\prime}(\phi)$ is evaluated at $\phi_{\text{min}}$,
i.e., $m_{\phi}^{2}=V_{eff}^{\prime\prime}(\phi)|_{\phi_{\text{min}}}$. (For a
potential of the form $V=M^{4+n}/\phi^{n}$ we have $m_{\phi}^{2}=\left[
n(n+1)M^{4+n}\phi^{-(n+2)}+(\beta\kappa)^{2}\bar{\rho}e^{\beta\kappa\phi
}\right]  _{\phi_{\text{min}}}$.) This leads to a large mass $m_{\phi}$ in
regions of high density and a small mass in regions of low density. Therefore
the range $m_{\phi}^{-1}$ of the chameleon scalar is very small in regions of
high density, but is much larger in regions of low density. It is for this
reason that, according to the chameleon model, earth-based gravity (short
ranged $\phi$) differs from space-based gravity (long ranged $\phi$), and why
chameleon effects are hidden on earth. [For example, Eq.(12) in
Ref.\citep{KW PRL} gives values for the range of $\phi$ in the atmosphere and
above the atmosphere (and on solar system scales) of $m_{\text{atm}}%
^{-1}\lesssim1$mm $-$1cm and $m_{\text{ss}}^{-1}\lesssim10-10^{4}$ AU, respectively.]

\bigskip

Since the chameleon's effects are hidden on earth, laboratory experiments
reveal no deviations from Newtonian gravity. [See\citep{KW PRD,KW PRL} for
estimates and arguments that laboratory tests of gravity are satisfied.]
Furthermore, a large body having a so-called \textquotedblleft thin
shell\textquotedblright\ will have a substantial chameleonic suppression in
comparison to that of a small body having a \textquotedblleft thick
shell\textquotedblright\ \citep{KW PRD,KW PRL}. For example, a large (thin
shelled) planet may behave differently from a small (thick shelled) satellite,
with the satellite showing an extra chameleonic acceleration not exhibited by
a planet. Consequently, the chameleon model successfully passes all existing
solar system constraints \cite[see Sec.VII]{KW PRD}.

\subsection{Remarks}

\textbf{(1)} There are now two contributions to the acceleration $\vec{a}$ of
a test mass, the Newtonian part, $\vec{a}_{N}=-\nabla\Phi$, and the scalar
chameleonic acceleration, $\vec{a}_{\phi}=-\nabla(\ln A)=-\beta\kappa
\nabla\phi$. So from (\ref{e6c}) (\textquotedblleft geodesic\textquotedblright%
\ equation), $\vec{a}=\vec{a}_{N}+\vec{a}_{\phi}$.

\bigskip

\textbf{(2)} The effective potential $V_{eff}(\phi)$ depends upon the local
matter density $\bar{\rho}$, and the result is that in regions of high
density, $\phi$ is small, $m_{\phi}$ is large, and the chameleon effect is
weak or undetectable, with the effects of the scalar being of very short
range. Then for a small $\phi$, $A(\phi)\approx1$ and in this case the
chameleonic acceleration $\vec{a}_{\phi}=-\nabla\left(  \ln A\right)
\approx0$ and $\vec{a}\approx\vec{a}_{N}=-\nabla\Phi$.

\bigskip

\textbf{(3) }However, in regions of low density, $m_{\phi}$ is very small, the
effects of the scalar are of long range, and the chameleon effect becomes
stronger. Therefore, for a satellite in earth orbit outside the earth and the
bulk of its atmosphere [taken to be roughly 10 km thick \citep{KW PRL}], the
chameleon effect can become stronger, with the possibility of a detectable
deviation from geodesic motion.

\section{Acceleration and Potential Fields}

\subsection{Effective gravitational potential and acceleration}

The chameleonic \textquotedblleft anomaly\textquotedblright\ would not be
apparent near the surface of the earth, but outside the earth's atmosphere it
could become detectable. This leads to a question of whether there is any
detectable discrepancy between gravitational fields and potentials determined
by earth-based measurements and space-based measurements from satellite
geodesy, both of which can possess high degrees of sensitivity. Presently, we
consider possible deviations from geodesic motion for earth satellites, such
as the LAGEOS satellite. If any measurable deviation from geodesic motion is
associated with the chameleon effect, it becomes possible to establish
approximate constraints on the chameleon model parameters.

From Eq.(\ref{e6c}) we have%
\end{subequations}
\begin{equation}
\nabla\cdot\vec{a}=-\nabla^{2}\Phi-\nabla^{2}(\ln A)=-\nabla^{2}\Phi
-\beta\kappa\nabla^{2}\phi\equiv-\nabla^{2}\Psi\label{3}%
\end{equation}

where the effective gravitational potential $\Psi(\vec{r})$ is%
\begin{equation}
\Psi=\Phi+\ln A(\phi)=\Phi+\beta\kappa\phi\label{4}%
\end{equation}

and the effective gravitational acceleration is%
\begin{equation}
\ \ \ \vec{a}=-\nabla\Psi=-\nabla\left(  \Phi+\delta\Phi\right)  \label{5}%
\end{equation}

with $\delta\Phi=\ln A=\beta\kappa\phi$. The 1st term $\nabla^{2}\Phi$ is
given by the 1st equation in (\ref{e6a}). Using this with (\ref{e6b}) we can
write
\begin{equation}
\nabla\cdot\vec{a}=-4\pi G\left\{  \bar{\rho}A-2V+\frac{2\beta}{\kappa}%
\nabla^{2}\phi\right\}  =-4\pi G\rho_{eff} \label{6}%
\end{equation}

where%
\begin{equation}
\rho_{eff}\equiv\bar{\rho}A-2V+\frac{2\beta}{\kappa}\nabla^{2}\phi\label{7}%
\end{equation}

We can therefore write a Poisson equation for $\Psi$,%
\begin{equation}
\nabla^{2}\Psi=4\pi G\rho_{eff}=4\pi G\left\{  \bar{\rho}A-2V+\frac{2\beta
}{\kappa}\nabla^{2}\phi\right\}  \label{8}%
\end{equation}

\subsection{Multipole expansion}

Let us apply this to a satellite orbiting earth outside the atmosphere, where
$r\gg r_{S}$ and we approximate $\bar{\rho}=0$. (The mass density is taken to
be the ambient density in our neighborhood of the galaxy, $\bar{\rho}%
_{G}\approx10^{-24}$ g/cm$^{3}$ \citep{KW PRL}, so that we take $\bar{\rho
}\approx0$. Also, from \citep{HF}, \citep{AM}, we have $\ln A=\beta\kappa
\phi\lesssim2\times10^{-12}\ll1$ so that $\ln A\ll1$. See, e.g., \citep{AM}
for a transcription of the results of \citep{HF} into our notation and
conventions.) Then the vacuum value of $\phi$ becomes large (with $\beta
\phi\ll1/\kappa$), $m_{\phi}$ becomes small (low curvature, an almost flat
chameleon potential $V(\phi)$), and $V(\phi)$ becomes very small.

\bigskip

Let us obtain solutions for $\Psi$ in terms of multipole moments which can be
obtained from measurements at the earth's surface, and multipole moments which
can be obtained from satellite measurements. Any difference between the two
sets of multipole measurements could indicate the presence of some type of
screened gravity effect.

\bigskip

We start with the formal solution to the Poisson equation (\ref{8}),%
\begin{equation}
\Psi(\vec{x})=-G\int_{V}\frac{\rho_{eff}(\vec{x}^{\prime})}{|\vec{x}-\vec
{x}^{\prime}|}d^{3}x^{\prime}\label{a1}%
\end{equation}

along with%
\begin{equation}
\frac{1}{|\vec{x}-\vec{x}^{\prime}|}=4\pi\sum_{l=0}^{\infty}\sum_{m=-l}%
^{l}\frac{1}{2l+1}\frac{r_{<}^{l}}{r_{>}^{l+1}}Y_{l}^{m\ast}(\theta^{\prime
},\varphi^{\prime})Y_{l}^{m}(\theta,\varphi) \label{a2}%
\end{equation}

where $\vec{x}^{\prime}$ is the source point, $\vec{x}$ is the field point,
and $r_{<}\ (r_{>})$ is the smaller (larger) of $|\vec{x}|$ and $|\vec
{x}^{\prime}|$. We consider our system to be the earth, with volume $V_{E}$,
surrounded by the atmosphere forming a thin shell of volume $V_{atm}$, along
with any chameleon field $\phi$ which may be present, and possibly
nonnegligble above the atmosphere. The chameleon field levels off away from
the earth, approaching an asymptotic value of $\phi_{\infty}$ \citep{KW PRL},
with $\nabla^{2}\phi\rightarrow0$.

\bigskip

\textbf{At earth's surface:}\ \ On the surface of the earth the chameleon
effect is absent, and the density interior to the earth's surface is $\rho
_{E}(\vec{x}^{\prime})$, i.e., just the ordinary earth density. The density
exterior to the earth's surface is the atmospheric density $\rho_{atm}$, along
with a chameleon contribution to the density, given by (\ref{7}), with
$\bar{\rho}=0$ (the chameleon effect is absent within the atmosphere, as
well). We note that $\rho_{eff}\rightarrow0$ as $|\vec{x}|\rightarrow\infty$.
Using (\ref{a1}) and (\ref{a2}) we have%
\begin{align}
\Psi_{l}^{m}(\vec{x})  &  =-4\pi G\int_{V_{E}}d^{3}x^{\prime}\left\{
\frac{\rho_{E}(\vec{x}^{\prime})}{2l+1}\frac{r^{\prime l}}{r^{l+1}}%
Y_{l}^{m\ast}(\theta^{\prime}\varphi^{\prime})Y_{l}^{m}(\theta,\varphi
)\right\} \nonumber\\
&  \ \ \ \ -4\pi G\int_{V_{\infty}}d^{3}x^{\prime}\left\{  \frac{\rho
_{eff}(\vec{x}^{\prime})}{2l+1}\frac{r^{l}}{r^{\prime l+1}}Y_{l}^{m\ast
}(\theta^{\prime}\varphi^{\prime})Y_{l}^{m}(\theta,\varphi)\right\}
\nonumber\\
&  =-4\pi G\left\{  \frac{A_{l}^{m}}{2l+1}\frac{Y_{l}^{m}(\theta,\varphi
)}{r^{l+1}}+\frac{I_{l}^{m}}{2l+1}r^{l}Y_{l}^{m}(\theta,\varphi)\right\}
,\ \ \ \ \ (r\approx r_{E}) \label{a3}%
\end{align}

where $r_{E}$ is the average radius of the earth, radii $r$ lie between the
earth's surface and the upper boundary of the atmosphere, $V_{\infty}$
includes all of space outside of the earth and atmosphere, the $\Psi_{l}^{m}$
are the multipole fields of $\Psi=\sum_{l,m}\Psi_{l}^{m}$, and the exterior
and interior multipole moments are given by%
\begin{equation}
A_{l}^{m}=\int_{V_{E}}d^{3}x^{\prime}\rho_{E}(\vec{x}^{\prime})r^{\prime
l}Y_{l}^{m\ast}(\theta^{\prime},\varphi^{\prime}) \label{a4}%
\end{equation}

and%
\begin{equation}
I_{l}^{m}=\int_{V_{\infty}}d^{3}x^{\prime}\frac{\rho_{eff}(\vec{x}^{\prime}%
)}{r^{\prime l+1}}Y_{l}^{m\ast}(\theta^{\prime},\varphi^{\prime}) \label{a5}%
\end{equation}

respectively. Furthermore, we neglect the contribution due to the interior
multipole moments $I_{l}^{m}$ by virtue of the comparative smallness of
$\rho_{eff}/r^{\prime l+1}$, or more specifically $(r^{l}|_{r_{E}})I_{l}%
^{m}\ll A_{l}^{m}/(r^{l+1}|_{r_{E}})$, with $\rho_{eff}\rightarrow0$ at points
distant from earth. We therefore have, approximately,%
\begin{equation}
\Psi(\vec{x})=\sum_{l=0}^{\infty}\sum_{m=-l}^{l}\Psi_{l}^{m}(\vec{x})=-4\pi
G\sum_{l=0}^{\infty}\sum_{m=-l}^{l}\frac{A_{l}^{m}}{2l+1}\frac{Y_{l}%
^{m}(\theta,\varphi)}{r^{l+1}},\ \ \ \ \ (r\approx r_{E}) \label{a6}%
\end{equation}

where the $A_{l}^{m}$ coefficients are determined by earth-based measurements.
This earth-based potential is just the Newtonian potential $\Phi$, i.e.,
$\Psi=\Phi$ near earth's surface.

\bigskip

\textbf{Above the atmosphere:}\ \ Above the atmosphere, but not at distances
too far from the earth, we again ignore interior multipole terms due to the
chameleon contribution to $\rho_{eff}$, which is very small compared to
$\rho_{E}$ and vanishes well away from earth, (or more specifically we assume
$(r^{l})I_{l}^{m}\ll A_{l}^{m}/(r^{l+1})$, for radial positions $r$ of the
satellite ) but at the position of a satellite in orbit above the atmosphere
we have the exterior moments%
\begin{equation}
B_{l}^{m}=\int_{Vsat}d^{3}x^{\prime}\rho_{eff}(\vec{x}^{\prime})r^{\prime
l}Y_{l}^{m\ast}(\theta^{\prime},\varphi^{\prime}) \label{a7}%
\end{equation}

where $V_{sat}$ is the volume of space interior to a surface on which \ the
satellite's orbit lies, where on this surface we assume that $\rho
_{eff}=-2V+\frac{2\beta}{\kappa}\nabla^{2}\phi$ may be small, but is not
assumed to vanish identically. We have that $\rho_{eff}\geq\rho_{E}$ within
the volume $V_{sat}$. From (\ref{7}), along with the expectation that
$|\nabla^{2}\phi|$ will maximize at some finite distance above the atmosphere
(see, for example, analytical and numerical solutions presented in
\citep{KW PRL}), we anticipate that $B_{l}^{m}-A_{l}^{m}\neq0$, and may be
measurable, if there does exist a chameleon effect. At points above the
atmosphere we therefore have%
\begin{equation}
\Psi(\vec{x})=\sum_{l=0}^{\infty}\sum_{m=-l}^{l}\Psi_{l}^{m}(\vec{x})=-4\pi
G\sum_{l=0}^{\infty}\sum_{m=-l}^{l}\frac{B_{l}^{m}}{2l+1}\frac{Y_{l}%
^{m}(\theta,\varphi)}{r^{l+1}},\ \ \ \ \ (r>r_{E}) \label{a8}%
\end{equation}

where the $B_{l}^{m}$ are determined by space-based satellite measurements.

\bigskip

\textbf{Acceleration:}\ \ The effective gravitational acceleration $\vec
{a}=\vec{a}_{N}+\vec{a}_{\phi}$ is given in terms of the gravitational
potential $\Psi$ by $\vec{a}=-\nabla\Psi$. We can therefore write the
acceleration, as measured by a satellite, as%
\begin{equation}
\vec{a}=-\nabla\Psi=-\sum_{l=0}^{\infty}\sum_{m=-l}^{l}\nabla\Psi_{l}%
^{m}(r,\theta,\varphi)=\sum_{l=0}^{\infty}\sum_{m=-l}^{l}\vec{a}_{l}%
^{m};\ \ \ \ \vec{a}_{l}^{m}=-\nabla\Psi_{l}^{m} \label{11}%
\end{equation}

where the $a_{i,l}^{\ m}$ are the multipole terms of the acceleration field
components $a_{i}$, ($i=r,\theta,\varphi$),
\begin{subequations}
\label{12}%
\begin{align}
a_{r,l}^{\ m}  &  =-\partial_{r}\Psi_{l}^{m}=-4\pi G\left[  B_{l}^{m}%
\frac{(l+1)}{r^{l+2}}Y_{l}^{m}(\theta,\varphi)\right] \label{12a}\\
a_{\theta,l}^{\ m}  &  =-\frac{1}{r}\partial_{\theta}\Psi_{l}^{m}=-4\pi
G\left[  -B_{l}^{m}\frac{1}{r^{l+2}}\partial_{\theta}Y_{l}^{m}(\theta
,\varphi)\right] \label{12b}\\
a_{\varphi,l}^{\ m}  &  =-\frac{1}{r\sin\theta}\partial_{\varphi}\Psi_{l}%
^{m}=-4\pi G\left[  -B_{l}^{m}\frac{1}{r^{l+2}}\frac{\partial_{\varphi}%
Y_{l}^{m}(\theta,\varphi)}{\sin\theta}\right]  \label{12c}%
\end{align}

For pure Newtonian gravity, characterized by a potential $\Phi$, i.e., for
points near the earth's surface where $\Psi\rightarrow\Phi$, or for points in
space for the case that there is no chameleon field and $\Psi=\Phi$, we would
have the same expressions for multipole terms with replacements $B_{l}%
^{m}\rightarrow A_{l}^{m}$.

\section{Comparison}

The potential $\Phi(r,\theta,\varphi)$ for pure Newtonian gravity is given by
a multipole expansion with multipole moments $A_{l}^{m}$, and the potential
$\Psi(r,\theta,\varphi)$ for chameleon gravity is given by (\ref{a8}). The
difference between each multipole term is
\end{subequations}
\begin{equation}
\delta\Phi_{l}^{m}\equiv\Psi_{l}^{m}-\Phi_{l}^{m}=-4\pi G(B_{l}^{m}-A_{l}%
^{m})\left[  \frac{1}{r^{l+1}}Y_{l}^{m}(\theta,\varphi)\right]  \label{13}%
\end{equation}

which would vanish identically if there were no chameleon field or chameleon
effect. However, if a detectable chameleon field does exist, it would give
rise to an \textquotedblleft anomalous\textquotedblright\ acceleration
$\vec{a}_{\phi}=\vec{a}-\vec{a}_{N}\equiv\delta\vec{a}$ with multipole terms%
\begin{equation}
\delta\vec{a}_{l}^{m}(r,\theta,\varphi)=-\nabla\delta\Phi_{l}^{m}=-(B_{l}%
^{m}-A_{l}^{m})(-4\pi G)\nabla\left[  \frac{1}{r^{l+1}}Y_{l}^{m}%
(\theta,\varphi)\right]  \label{14}%
\end{equation}

However, the acceleration field $\vec{a}$ measured near the earth would not
exhibit an anomalous acceleration, i.e., $\delta\vec{a}=0$, while the
anomalous acceleration predicted by the chameleon effect that could be
detected by a satellite in orbit would be nonzero. From (\ref{13}) and
(\ref{14}) we have a relative correction to the Newtonian fields, due to the
chameleon effect,%
\begin{equation}
\frac{\delta\Phi_{l}^{m}(\vec{r})}{\Phi_{l}^{m}(\vec{r})}=\left(  \frac
{B_{l}^{m}}{A_{l}^{m}}-1\right)  ,\ \ \ \ \ \ \frac{|\delta\vec{a}_{l}%
^{m}(\vec{r})|}{a_{N,l}^{\ \ m}(\vec{r})}=\left(  \frac{B_{l}^{m}}{A_{l}^{m}%
}-1\right)  \label{15}%
\end{equation}

The $B_{l}^{m}$ coefficients are determined from satellite measurements of
$\vec{a}$, and the $A_{l}^{m}$ coefficients are determined from the
ground-based measurements of $\vec{a},$ where the chameleon effect disappears.

\bigskip

The acceleration $\vec{a}$ and its \textquotedblleft
anomalous\textquotedblright\ part $\delta\vec{a}$ are dominated by the
monopole and quadrupole terms. (The $l=1$ dipole term vanishes, as we take the
coordinate origin to coincide with the center of mass.) The monopole radial
acceleration \textquotedblleft anomaly\textquotedblright\ is
\begin{equation}
\frac{|(\delta a_{r})_{0}^{0}|}{(a_{N})_{0}^{0}}=\left(  \frac{B_{0}^{0}%
}{A_{0}^{0}}-1\right)  \label{16}%
\end{equation}

For a uniform sphere of mass having only a monopole Newtonian field and no
chameleon anomaly, i.e., $A_{l}^{m}=B_{l}^{m}$, the gravitational field is
$a_{N,r}=-GM/r^{2}=-4\pi GA_{0}^{0}Y_{0}^{0}/r^{2}$ (see Eq.(\ref{12a}), for
example). In this case, the \textquotedblleft geoid\textquotedblright\ would
be a sphere at the sphere's surface, and we would identify $A_{0}^{0}%
=M/\sqrt{4\pi}$. If there is a chameleonic anomaly, with $B_{0}^{0}\neq
A_{0}^{0}$, then the \textquotedblleft anomaly\textquotedblright\ may take an
appearance of a slightly modified gravitational parameter, $(GM)_{eff}%
=GM+\delta(GM)$, with $\delta(GM)/(GM)\propto(B_{0}^{0}/A_{0}^{0}-1)$.

\bigskip

Although a gravitational anomaly resulting from differences in space-based and
ground-based measurements of the gravitational field may be rather small, its
existence could give some credence to the idea of some theory of modified
gravity with a screening mechanism, such as the chameleon model of gravitation.

\section{Numerical Estimates: Monopole Approximation}

For a simple example, we now make a monopole approximation for the earth's
gravitational field, treating the earth as a uniform sphere of radius $R$ and
mass $M=\int\rho_{E}d^{3}x$. The magnitude of the Newtonian gravitational
field at a distance $r$ from the center of the earth is
\begin{equation}
a_{N}(r)=\frac{4\pi GA_{0}^{0}Y_{0}^{0}}{r^{2}}=\frac{GM}{r^{2}} \label{17}%
\end{equation}

where $A_{0}^{0}=M/\sqrt{4\pi}$ is the monopole ($l=0,m=0$) moment
contribution to the gravitational field. Both the Newtonian acceleration
$\vec{a}_{N}$ and the chameleonic acceleration $\vec{a}_{\phi}$ are directed
radially inward \citep{AM}. To get a numerical estimate for $\delta
a(r)/a_{N}(r)$, we use results that were reported in Ref.\citep{AM}, where the
anomalous acceleration $\delta a$ is identified with the chameleonic
acceleration $a_{\phi}$. For the present case of an earth satellite we have
the result%
\begin{equation}
\frac{|\delta a_{r}|}{a_{N}}=\left(  \frac{B_{0}^{0}}{A_{0}^{0}}-1\right)
=\frac{|a_{\phi}|}{a_{N}}=6\beta^{2}\Delta_{E} \label{18}%
\end{equation}

where $\beta$ is a constant denoting the strength of the coupling of the
chameleon field to matter, and $\Delta_{E}$ is the \textquotedblleft thin
shell\textquotedblright\ factor for the earth (the gravitational
\textquotedblleft source\textquotedblright). Here, we are taking $\delta
a_{r}$ to be an inward radial acceleration, which should appear to be a small
radial acceleration of the satellite, in excess of the Newtonian acceleration.
On the other hand, if it is determined that there is no inward radial
component of $\delta\vec{a}$, then we must conclude that there is no
chameleonic acceleration, i.e., $\beta=0$. Also note that the ratio $B_{0}%
^{0}/A_{0}^{0}=(GM)_{\text{sat}}/(GM)_{E}$ where $(GM)_{\text{sat}}$ is
measured by satellite from space and $(GM)_{E}$ is measured near the earth's surface.

\bigskip

Several estimates or bounds can be placed on the \textquotedblleft
anomalous\textquotedblright\ acceleration $\delta a_{r}$.

\subsection{Estimated Bounds}

\textbf{Khoury-Weltman bound:}\ \ In Refs. \citep{KW
PRD} and \citep{KW PRL} the value of $\beta$ is taken to be on the order of
unity, $\beta\approx1$, and the thin shell factor for the earth is estimated
to be $\Delta_{E}<10^{-7}$. With these parameters adopted by KW, we have a
rough estimate for an upper bound of $|\delta a_{r}|/a_{N}$ given by%
\begin{equation}
\left(  \frac{|\delta a_{r}|}{a_{N}}\right)  _{KW}\lesssim6\beta^{2}%
\times10^{-7}\approx6\times10^{-7}\ \ (\beta\approx1) \label{19}%
\end{equation}

\bigskip

\textbf{Cassini bound:}\ \ On the other hand, we can appeal to an upper limit
based upon solar system constraints obtained by the Cassini mission, where
limits were established \citep{Bertotti} on the parameterized Post-Newtonian
(PPN) parameter $\gamma$,%
\begin{equation}
\gamma-1=(2.1\pm2.3)\times10^{-5} \label{20}%
\end{equation}

Hees and Fuzfa \citep{HF} have used this solar system-based constraint to
obtain an upper bound on chameleon parameters, without making assumptions
regarding the values of $\beta$ or the thin shell factors. This analysis was
incorporated in Ref. \citep{AM} to obtain (see Ref. \citep{AM} for details)%
\begin{equation}
\beta^{2}\Delta_{S}\lesssim3.3\times10^{-7} \label{21}%
\end{equation}

where $\Delta_{S}$ is the thin shell factor for the sun. The relation between
$\Delta_{S}$ and $\Delta_{E}$ is given by \citep{AM}%
\begin{equation}
\Delta_{E}\approx\left(  \frac{1}{3}\times10^{4}\right)  \Delta_{S} \label{22}%
\end{equation}

giving an upper bound on $\beta^{2}\Delta_{E}$ based upon the Cassini
constraint%
\begin{equation}
\left(  \beta^{2}\Delta_{E}\right)  _{C}\lesssim10^{-3} \label{23}%
\end{equation}

From (\ref{18}) we then have a Cassini-based estimate%
\begin{equation}
\left(  \frac{|\delta a_{r}|}{a_{N}}\right)  _{C}=6\beta^{2}\Delta_{E}%
\lesssim6\times10^{-3} \label{24}%
\end{equation}

\bigskip

\textbf{LAGEOS bound:}\ \ Consider now a fifth force Yukawa addition to the
geopotential,%
\begin{equation}
\Psi(r)=\Phi(r)+\delta\Phi(r)=-\frac{GM}{r}\left(  1+\alpha e^{-r/\lambda
}\right)  =\Phi(1+\alpha e^{-r/\lambda}) \label{25}%
\end{equation}

where $\alpha$ represents the coupling of the fifth force, which we will here
assume to be universal, and $\lambda$ represents the range of the Yukawa
potential. We are interested in the effect of $\delta\Phi=-\alpha\Phi
e^{-r/\lambda}$ on a satellite where the ambient mass density is negligible,
and $m_{\phi}\approx0$ and $\lambda\rightarrow\infty$. In this case we have
$\delta\Phi=\alpha\Phi$ and $\delta\Phi/\Phi=\alpha.$ Identifying $\delta
\Phi/\Phi=|\delta a_{r}|/a_{N}$ [see Eq.(\ref{15})] and using (\ref{18}), we
have%
\begin{equation}
\frac{\delta\Phi}{\Phi}=\frac{|\delta a_{r}|}{a_{N}}=\left(  \frac{B_{0}^{0}%
}{A_{0}^{0}}-1\right)  =6\beta^{2}\Delta_{E}=|\alpha| \label{26}%
\end{equation}

The upper bound on $\alpha$, as determined by the LAGEOS satellite, is quoted
to be \citep{Iorio,Ciufolini95}
\begin{equation}
|\alpha|=\left(  \frac{|\delta a_{r}|}{a_{N}}\right)  _{L}=6\beta^{2}%
\Delta_{E}<10^{-5}-10^{-8} \label{alpha}%
\end{equation}

Of course, for this fifth force to be associated with the chameleon effect,
$\alpha$ should be positive.

\bigskip

To summarize, the deviation from Newtonian acceleration, as well as the
fractional change in the monopole moment is, from (\ref{16}), (\ref{18}),
(\ref{19}), (\ref{24}), and (\ref{alpha}), estimated as%
\begin{equation}
\frac{|\delta a_{r}|}{a_{N}}=\left(  \frac{B_{0}^{0}}{A_{0}^{0}}-1\right)
=6\beta^{2}\Delta_{E}\lesssim\left\{
\begin{array}
[c]{cc}%
6\times10^{-7}, & \text{(KW, }\beta\approx1\text{)}\\
6\times10^{-3}, & \text{(Cassini)}\\
10^{-5}-10^{-8}, & \text{(LAGEOS)}%
\end{array}
\right\}  \label{28}%
\end{equation}

where use has been made of (\ref{19}), using the original fiducial KW
parameters with $\beta\approx1$, along with (\ref{23}), and (\ref{26}). The
discrepancy between earth-based and space-based monopole moments may be quite
small, but the measurement of a deviation from geodesic motion for the LAGEOS
satellite is fairly sensitive, and does not assume any particular values for
$\beta$ or $\Delta_{E}$. The Cassini-based constraint allows much more
freedom, but again, is based upon fewer assumptions than the KW estimate. The
LAGEOS bound is seen to provide more restriction than the Cassini bound, and
does not assume values for $\beta$ or $\Delta_{E}$.

\bigskip

\textbf{Note:}\ \ A cautionary note is in order here. The formalism presented
here, and the subsequent constraints obtained, have been admittedly
oversimplified. Effects on satellite acceleration beyond a simple Newtonian
acceleration, along with a possible chameleonic correction, have been ignored.
Such effects can include gravitational forces from the sun, moon, and other
planets, solar radiation pressure, and general relativistic effects (which for
a scalar-tensor theory might be accommodated by appropriate Eddington
parameters). (See, for example, \citep{Combrinck} and \citep{Damour} and
references therein.) A meaningful comparison of theoretical predictions and
satellite data is therefore expected to require delicate expertise, and is
well beyond the scope and intent of the basic ideas presented here.

\section{Summary}

A theoretical framework has been proposed by which a gravitational anomaly due
to a scalar \textquotedblleft chameleon\textquotedblright\ field might be
detected by comparing measurements of the gravitational field strength made at
the earth's surface and from a satellite in earth orbit. The chameleon model
allows such an \textquotedblleft anomalous\textquotedblright\ acceleration to
appear at points far from a matter source (satellite in orbit), but the
anomaly does not become manifest at points near a matter source (earth's surface).

\bigskip

Gravitational field measurements made via satellite can be compared with
earth-based measurements where the chameleon effect is absent. The presence of
a difference in the two sets of measurements could indicate the existence of a
chameleonic gravitational correction. On the other hand, clear evidence of an
absence of anomaly could severely constrain the chameleon model of gravitation.

\bigskip

Estimates based upon a simple monopole approximation indicate that the
relative difference between earth-based multipole moments $A_{l}^{m}$ and
space-based multipole moments $B_{l}^{m}$ is expected to be very small, with
an estimate for a difference in monopole moments given by (\ref{28}) if the
difference is due solely to the chameleon effect. If, for instance, the
acceleration and coefficients $A_{l}^{m}$ are obtained on earth's surface,
near the geoid, at some position $(R,\theta,\varphi)$ and the corresponding
acceleration and coefficients $B_{l}^{m}$ are determined by a satellite in
orbit at position $(r,\theta,\varphi)$ (at approximately the same time to
eliminate temporal differences), they can be compared. Since the multipole
moments are obtained from acceleration measurements, and not computed from the
formal expressions involving $\rho_{eff}$, one need not be concerned with any
dependence upon spatial uncertainties in the density $\bar{\rho}$. However,
given the fact that the higher multipole moments are typically smaller than
the lower ones, it is hopeful that sufficiently sensitive measurements of the
difference $(B_{l}^{m}-A_{l}^{m})$ could indicate whether space-based gravity
is indeed different from earth-based gravity.

\bigskip

\ \

\end{document}